\documentclass[aps,pra,superscriptaddress,twocolumn]{revtex4}
\usepackage{amsmath}
\usepackage{graphicx}	
\usepackage{bm} 
\usepackage{color}
\usepackage{setspace}
\usepackage{placeins}

\begin{document}
\title{Optomechanics with a position-modulated Kerr-type nonlinear coupling}
\date{\today}
\author{M.~Mikkelsen}
\email[]{mathias.mikkelsen@oist.jp}
\affiliation{Quantum Systems Unit, OIST Graduate University, Onna, Okinawa 904-0495, Japan}
\author{T.~Fogarty}
\affiliation{Quantum Systems Unit, OIST Graduate University, Onna, Okinawa 904-0495, Japan}
\author{J.~Twamley}
\affiliation{Centre for Engineered Quantum Systems, Department of Physics and Astronomy, Macquarie University, NSW 2109, Australia}
\author{Th.~Busch}
\affiliation{Quantum Systems Unit, OIST Graduate University, Onna, Okinawa 904-0495, Japan}

\begin{abstract} 
Cavity optomechanics has proven to be a field of research rich with possibilities for studying motional cooling, squeezing, quantum entanglement and metrology in solid state systems. While to date most studies have focused on the modulation of the cavity frequency by the moving element, the emergence of new materials will soon allow us to explore the influences of nonlinear optical effects. 
We therefore study in this work the effects due to a nonlinear position-modulated self-Kerr interaction 
and find that 
this leads to an effective coupling that scales with the square of the photon number, meaning that significant effects appear even for very small nonlinearities. This strong effective coupling can lead to lower powers required for motional cooling 
and the appearance of multistability in certain regimes. 
\end{abstract}

\maketitle

\section{Introduction}
Cavity optomechanics studies radiation-pressure-induced coherent photon-phonon
interactions and has in recent years led to a rich trove of novel phenomena \cite{RevModPhys.86.1391}, including phonon cooling \cite{Teufel2011, Chan2011}, optomechanically induced transparency~\cite{Weis1520, Safavi2011}, and mechanical squeezing \cite{Brooks2012,Purdy2013,Safavi2013, Wollman2015, Pirkkalainen2015}. It has also found use in other areas such as the generation of hybrid entanglement \cite{Wang2013, Palomaki2013, Vivoli2016, Riedinger2016}, quantum computation \cite{Schmidt2012, Rips2013, Houhou2015}, and precision sensing \cite{Forstner2012, Bagci2014, Wu2014}. 


Nonlinear effects in such systems are currently a frontier research topic, and it is well understood that optomechanical systems 
with a linear optomechanical interaction, $\hat{H}_\text{int}\sim \hat{a}^\dagger \hat{a}\hat{q}$, 
can yield  {\it mechanical} Kerr-type nonlinear optical responses \cite{Aldana2013}.  Furthermore, optomechanical systems whose interactions are higher order in the position of the mechanical element, e.g. $\hat{H}_\text{int}\sim \hat{a}^\dagger \hat{a}\,\hat{q}^2$ have been studied \cite{Vanner2011, Paraiso2015, Brawley2016} and shown to allow for new ways of coherent control of matter. Additionally a cross-Kerr interaction between the optical and mechanical modes, i.e. $\hat{H}_\text{int}\sim \hat{a}^\dagger \hat{a} \hat{b}^\dagger \hat{b}$,  was considered in \cite{Xiong2016} and was shown to stabilize the bistable solutions as well as leading to tristable solutions in certain parameter regimes. In this work we will extend the range of nonlinear phenomena in cavity optomechanics by considering a situation where a nonlinear medium is present in the cavity. This situation has recently attracted some attention where a $\chi^{(3)}$ material in the cavity was considered \cite{Kumar2010, Shahidani2013}, and mechanical resonators with large optical nonlinearities are currently under development \cite{Lu2014, Lu2015}. Another route to generate large optical nonlinearities is via coherent processes in atoms and the $N$-system of Schmidt and Immamoglu \cite{Schmidt1996} is a well-known example. In our work, rather than considering a stationary optical nonlinearity, we will focus on the situation where the cavity's optical nonlinearity is directly modulated by the mechanical position, e.g. $\hat{H}_\text{int}\sim \hat{a}^{\dagger 2}\hat{a}^2\hat{q}$. Such an interaction can be engineered in superconducting systems, cavity polaritonic systems and atom-optical systems.  For example, a giant-self-Kerr microwave optical nonlinearity was recently shown to be possible in a superconducting coplanar resonator via a capacitive coupling between two Cooper-pair boxes \cite{Rebic2009}. By utilizing an electromechanical capacitor (such as in Ref. \cite{Teufel2011a}), a giant-Kerr mechanical modulation can then be straightforwardly engineered. Cavity polaritons, where light strongly couples to excitons in a quantum well or dot semiconductor hetrostructure, have also been studied for their use in optomechanics \cite{Kyriienko2014}, and ultrastrong optomechanical couplings have been described \cite{Jusserand2015}. The large interactions present in cavity polaritons are good candidates for optomechanical Kerr modulations to be expected \cite{Bobrovska2016}. Strong coupling between light and matter on the scale of individual atoms has recently been reported for atoms trapped near fiber tapers \cite{Kato2015, Polzik2016}, and the generation of giant-Kerr optical nonlinearities using electromagnetically induced transparency techniques of $N$-systems (similar to Ref. \cite{Kumar2015}),  would also lead to a position dependent self-Kerr interaction. 

In this paper we will describe the general situation of nonlinear optical cavity optomechanics and include both, the usual optomechanical coupling and the Kerr-type coupling. For this we will first derive the relevant Langevin equations and calculate the classical steady-state solutions. We will then study the quantum fluctuations about these and the spectrum of the fluctuations. Using a numerical analysis we will discuss how the inclusion of the Kerr nonlinear interaction alters the cooling rates and other key figures of merit and end by analyzing the appearance of optical multistability and the situation of position dependent absorption.

\section{Theoretical  framework and analytic results}
In the following we will first outline the model and the theoretical framework we use. From this we will derive the analytic equations that govern the physical properties of the system.   

\subsection{Basic model}
The system we consider can be described by an extension of standard optomechanical setups, such as the membrane-in-the-middle model \cite{Bhattacharya2008, Thompson2008, Biancofiore2011}, where the membrane is modeled as a harmonic oscillator with a single mechanical frequency $\Omega_m$ and nondimensional position and momentum operators $(\hat{q},\hat{p})$. The cavity field is modeled as a harmonic oscillator and described by the bosonic creation and annihilation operators, $ \hat{a}^{\dagger}$ and $ \hat{a}$, which create photons with the resonant frequency $\omega_0$. The membrane and the cavity interact via radiation pressure, which pushes the membrane and therefore changes the resonance frequency, making it position dependent $\omega(\hat{q})$. Since the harmonic oscillator is described by $\hbar \omega(\hat{q}) \hat{a}^\dagger \hat{a}$, this gives rise to a position-dependent interaction which is linear in the optical-field operators.  While it is possible to perform the analytic treatment with no assumptions of  the form of $\omega(\hat{q})$ \cite{Biancofiore2011}, we will consider the case of  
\begin{equation}
\omega(\hat{q}) \approx \omega(0)+\frac{d\omega(\hat{q})}{d\hat{q}}\bigg|_{\hat{q}=0} \hat{q} = \omega_0-g_\text{L} \hat{q}
\label{eq:nonlinearcoupling}
\end{equation}
from the beginning as this clarifies the relation between the two types of interaction we will consider. Within the linearity assumption the optomechanical interaction is then described as $\hat{H}_\text{L,int}=-\hbar \hat{a}^{\dagger} \hat{a} g_\text{L}  \hat{q}$.  

The second type of interaction we consider is nonlinear in the optical field operators, which corresponds to a simultaneous two-photon process that can be facilitated by a $\chi^{(3)}$ material and is described by the term $H_{\text{Kerr}}= \eta \hbar \hat{a}^{\dagger} \hat{a}^{\dagger} \hat{a}\hat{a}$. In order to obtain a position-dependent interaction with a moving membrane the nonlinear coefficient has to become dependent on its position $\eta(\hat{q})$. If this can be engineered, then the nonlinear coefficient will be given similarly to the linear coefficient as 
\begin{equation}
\eta(\hat{q}) \approx \eta(0)+\frac{d\eta(\hat{q})}{d\hat{q}}\bigg|_{\hat{q}=0} \hat{q} = \eta_0-g_\text{NL} \hat{q}.
\end{equation}
Throughout the paper we will generally use the  dimensionless position-coordinate $\hat{q}$, but in order to derive the values of these coefficients for a specific physical system, it is easier to use the dimensional position coordinate $\hat{x}$, which is related to the nondimensional coordinate by some characteristic length scale $x_0$ as $\hat{q} = \hat{x}/x_0$. Correspondingly, one can relate coupling strengths for the nondimensional coordinates to those of the dimensional coordinates by $g_\text{NL}=x_0 G_\text{NL},g_\text{L}=x_0 G_\text{L}$.  The characteristic length scale in optomechanics is generally the zero-point motion of the mechanical element $x_\text{zp}$ \cite{RevModPhys.86.1391}.  
\subsubsection*{Physical implementation in an optical cavity}
A model where the entire space between the two mirrors (where one of them can move) of a cavity is filled by a  $\chi^{(3)}$ medium giving rise to a term $\hat{H}_{\text{Kerr}}$ was recently considered  in \cite{Kumar2010}. We note that this model leads to a position dependence of $\eta$ as 
\begin{equation}
\eta = \frac{3 \hbar \omega^2 \text{Re}[\chi^{(3)}]}{2 \epsilon_0 V_c},
\end{equation} 
and both, the resonant frequency $\omega(\hat{x})=\omega_0-G_\text{L} \hat{x}$ and the cavity volume $V_c(\hat{x})=(L_0+\hat{x})V_{c,0}/L_0$, depend on the position coordinate of the end mirror. Here $\epsilon_0$ is the vacuum permittivity,  while $L_0$ is the length of the cavity and $V_{c,0}$ is the cavity volume, both in the absence of coupling. If we evaluate the first-order derivative at $\hat{x}=0$ and use the form of the linear coupling in this setup $G_{\text{L}}=\omega_0/L_0$, then we find that  $G_\text{NL}=-3 \frac{\eta_0}{L_0}$. This means that a relatively big nonlinear coupling can be achieved only when the cavity length $L_0$ is small. Since  $x_\text{zp} \ll L_0$ always holds, however,  the interaction term $\hat{H}_{\text{NL,int}}=-\hbar \hat{a}^{\dagger} \hat{a}^{\dagger} \hat{a}\hat{a} g_\text{NL}  \hat{q}$ will be small compared to $\hat{H}_{\text{Kerr}}$. The photon blockade effects arising from the $\hat{H}_{\text{Kerr}}$ term which was the topic of the investigation in Ref. \cite{Kumar2010} would therefore obscure the physics of interest in this work. 
\subsubsection*{Physical implementation in a microwave cavity}
In order to overcome this issue and to obtain a variable strong nonlinear interaction we turn to a different setup in which $\eta$ is more tunable. It has been shown that a giant-self-Kerr microwave optical nonlinearity is possible in a superconducting coplanar resonator via a capacitive coupling between two Cooper-pair boxes \cite{Rebic2009}. While the details are too involved to present here (see Appendix), such a setup allows for a nonlinear coupling that depends on the mutual capacitance between the two Cooper-pair boxes. As the mutual capacitance $C_m$ depends on the distance between the plates, $C_m$ can be made dependent on $\hat{x}$ by coupling the motion of one of the Cooper-pair boxes to the physical motion of a membrane. By manipulating the parameters within physically realistic constraints, nonlinear couplings $g_\text{NL}$ of similar size to the typical linear couplings $g_\text{L}$ achievable in microwave cavities are obtainable ($g_\text{L},g_\text{NL} \sim  \text{a few kHz}$). The same caveat as in the optical case is still present, but due to the tunability of the artificial molecules it is possible to place a second molecule inside the cavity which does not couple to the position and has the same magnitude as $\eta_0$ but the opposite sign. See Appendix for more details and a discussion of the relevant physical parameters. As it is possible to engineer situations where $\eta_0 =0 $, we will ignore the constant Kerr-term in our Hamiltonian, as it will generally lead to photon blockade which diminishes the effects of the nonlinear interaction that we want to investigate. Instead we consider just the position-modulated nonlinear term, described by the interaction Hamiltonian  $\hat{H}_{\text{NL,int}}$.

Finally the light field is produced by an input laser which has a frequency $\omega_L$ and an energy $E$. The full Hamiltonian in the frame rotating at the driving laser frequency $\omega_L$ is then given by 
\begin{align}
& \hat{H}=\hbar \Delta \hat{a}^\dagger \hat{a} + \frac{\hbar \Omega_m}{2}(\hat{p}^2+\hat{q}^2) \nonumber \\
& +i \hbar E(\hat{a}^\dagger - \hat{a}) -\hbar \hat{a}^{\dagger} \hat{a} g_\text{L}  \hat{q}  -\hbar \hat{a}^{\dagger} \hat{a}^{\dagger} \hat{a}\hat{a} g_\text{NL}  \hat{q}.
\label{Eq:basichamiltonian}
\end{align}
Here the first term corresponds to the optical harmonic oscillator with detuning $\Delta=\omega_0-\omega_L$, while the second term corresponds to the single-frequency $\Omega_m$ mechanical oscillator and the third term corresponds to the the input laser field of strength $E$. The last two terms account for the linear and nonlinear interactions as described above. 

\subsection{Quantum Langevin formalism}
The optomechanical setup we are considering is an open system and interactions with the environment in the form of photon losses, mechanical dissipation, and noise have to be taken into account. We do this by utilizing the quantum Langevin formalism
and consider photon loss associated with the end mirrors of the cavity $\kappa_0$ and photon loss associated with a moving element, such as a membrane $\kappa_1(\hat{q})$.  While experimental evidence suggest that the dependence of  $\kappa_1(\hat{q})$ on the position is usually small \cite{Karuza2012}, we take it into account to get the most general picture. 
The losses have associated noise operators  $\hat{a}_0^{in}$ and $\hat{a}_1^{in}$, which have the correlation relation \cite{Biancofiore2011}
\begin{equation}
\langle \hat{a}_j^{in}(t)  (\hat{a}_j^{in})^\dagger(t') \rangle = \delta(t-t') . 
\end{equation}
In addition to the losses associated with the optical field, mechanical dissipation of the membrane $\gamma_m$ must be considered. For this the associated noise operator has the correlation relation 
\begin{align}
\langle \xi(t) \xi(t') \rangle &= \frac{\gamma_m}{\Omega_m} \int \frac{d \nu}{2\pi} e^{-i\nu(t-t')}\nu\left[1+\coth\left(\frac{\hbar\nu}{2 k_b T_0}\right)\right], \nonumber  \\
& \approx \gamma_m\left[(2 n_0+1)\delta(t-t)+i\frac{\delta'(t-t')}{\Omega_m}\right],
\end{align}
where $\nu$ is the Fourier space frequency and
\begin{equation}
n_0 = \frac{1}{e^{\hbar \Omega_m /k_B T_0}-1},
\end{equation}
is the mean phonon number at temperature $T_0$ and $\delta'(t-t')$ is the derivative of the Dirac-delta function. Adding these to the Heisenberg equations for the Hamiltonian (\ref{Eq:basichamiltonian}) we arrive at the quantum Langevin equations (QLE)
\begin{align}
& \dot{\hat{q}} = \Omega_m \hat{p} ,\\
& \dot{\hat{p}} =-\Omega_m \hat{q}+g_\text{NL}\hat{a}^{\dagger}\hat{a}^{\dagger} \hat{a}\hat{a}+g_\text{L} \hat{a}^\dagger \hat{a} \nonumber \\
& -\gamma_m \hat{p} +\hat{\xi}-i\frac{\partial_q\kappa_1(\hat{q})}{\sqrt{2\kappa_1(\hat{q})}}\left( \hat{a}^\dagger \hat{a}^{in}_1-\hat{a} (\hat{a}^{in}_1)^\dagger \right), \\
& \dot{\hat{a}}=-i(\Delta-g_\text{L}\hat{q}-2g_\text{NL}\hat{q} \hat{a}^\dagger \hat{a})\hat{a} \nonumber \\
& +E-[\kappa_0+\kappa_1(\hat{q})]\hat{a}+\sqrt{2\kappa_0}\hat{a}_0^{in}+\sqrt{2\kappa_1}\hat{a}_1^{in}.
\end{align}
The $\partial_q\kappa_1(\hat{q})$ term is an effective noise term arising from the absorption in the membrane, as long as it is not assumed constant \cite{Biancofiore2011}. Solving this full set of quantum-mechanical equations is not a trivial task and we therefore employ a semiclassical approximation where we assume the physical operators can be expressed as a classical  average  $\{q,p,\alpha\}$ plus some small quantum fluctuation $\{\delta \hat{q} ,\delta \hat{p},\delta \hat{a} \}$
\begin{equation}
\hat{q} = q+\delta \hat{q}  \quad \quad , \quad \quad \hat{p} = p+\delta \hat{p} \quad \quad , \quad \quad \hat{a} = \alpha+\delta \hat{a} .
\end{equation}
This assumption only holds when the classical part is large, which is the situations experiments are commonly in. In the next section we will look at various aspects of the solutions for the system.
\subsection{Classical steady-state solutions}
To determine the classical steady-state solutions 
we consider the quantum fluctuations to be small compared to the classical c-numbers, so that any terms containing them can be neglected. 
Setting all derivatives in the QLE to zero then leads to
\begin{align}
q_s &=(g_\text{L}+g_\text{NL}|\alpha_s|^2)\frac{|\alpha_s|^2}{\Omega_m}
\label{eq:steadystateq}\\
\alpha_s&=\frac{E}{\kappa(q_s)+i(\Delta-g_\text{L}q_s-2g_\text{NL}q_s|\alpha_s|^2)},
\end{align}
where $\kappa(q_s)=\kappa_0+\kappa_1(q_s)$. Assuming that $\kappa_1(q_s)=\kappa_\text{L} q_s$  these two coupled equations can be expressed as a single seventh-order polynomial for the mean photon number $n_s=|\alpha_s|^2$ of the form
\begin{align}
\frac{4 g_\text{NL}^4}{\Omega_m^2}& n_s^7+\frac{12g_\text{NL}^3 g_\text{L}}{\Omega_m^2}n_s^6+\frac{g_\text{NL}^2}{\Omega_m^2}(13g_\text{L}^2+\kappa_\text{L}^2)n_s^5\nonumber\\
&+\frac{g_\text{NL}}{\Omega_m^2}(6 g_\text{L}^3+2g_\text{L} \kappa_\text{L}^2-4g_\text{NL} \Delta \Omega_m)n_s^4 \nonumber\\
&+\left(\frac{g_\text{L}^2(g_\text{L}^2+\kappa_\text{L}^2)}{\Omega_m^2}-6\frac{g_\text{NL} g_\text{L} \Delta}{\Omega_m}+\frac{2 g_\text{NL} \kappa_0 \kappa_\text{L}}{\Omega_m}  \right)n_s^3\nonumber\\
&+\frac{2 g_\text{L}}{\Omega_m}( \kappa_0 \kappa_\text{L}-\Delta g_\text{L})n_s^2+(\Delta^2+\kappa_0^2)n_s-E^2 = 0.
\label{eq:nonlinearsteadystate}
\end{align}
Even though this seventh-order polynomial equation has in principle seven roots, all complex solutions can be discarded as the mean photon number has to be real. The steady-state position of the membrane can then be found by inserting the resulting solution for $n_s$ into expression (\ref{eq:steadystateq}). We note that doing a similar analysis for the linear system leads to a third-order polynomial only. 

\subsection{Linearised quantum Langevin equations: Effect of quantum fluctuations}
To determine the stability of the steady-state solutions and to calculate physical values that depend solely on the quantum fluctuations, such as the temperature of the membrane, we will in this section look at the effects of the fluctuation terms.  For this we use the quadratures of the electric field
$\hat{\delta X} = (\delta \hat{a}+\delta \hat{a}^\dagger)/\sqrt{2}$ , $\hat{\delta Y} = (\delta \hat{a}-\delta \hat{a}^\dagger)/\sqrt{2}i$, $\hat{X}^{in}_j = (\hat{a}_j^{in}+\hat{a}_j^{in \dagger})/\sqrt{2}$ and $\hat{Y}^{in}_j = (\hat{a}_j^{in}-\hat{a}_j^{in \dagger})/\sqrt{2}i$  and insert the steady-state solution plus the quantum-fluctuations into the QLEs. We are still considering the quantum fluctuations and the noise operators to be small and therefore only keep terms up to first order in these operators. This leads to the linearized quantum Langevin equations (LQLE) of the form
\begin{align}
&\dot{\delta \hat{q}} = \Omega_m \delta \hat{p}, \\
&\dot{\delta \hat{p}}=-\Omega_m  \delta \hat{q}-\gamma_m \delta \hat{p}+G \delta \hat{X}+\hat{\xi}+\frac{\Gamma}{2\sqrt{\kappa_1(q_s)}}\hat{Y}^{in}_1,\\
&\dot{\delta \hat{X}}=-\kappa(q_s)\delta \hat{X}+(\Delta(q_s)-2 g_\text{NL} q_s)\delta \hat{Y} \nonumber \\
&-\Gamma \delta \hat{q}+\sqrt{2\kappa_0}\hat{X}_0^{in}+\sqrt{2\kappa_1(q_s)}\hat{X}_1^{in}, \\
&\dot{\delta \hat{Y}}=-\kappa(q_s)\delta \hat{Y}-(\Delta(q_s)-6 g_\text{NL} q_s)\delta \hat{X} \nonumber \\
&+G \delta \hat{q}+ \sqrt{2\kappa_0}\hat{Y}_0^{in}+\sqrt{2\kappa_1(q_s)}\hat{Y}_1^{in},
\end{align}
where we have defined the effective loss rate due to the position-dependent $\kappa_1(\hat{q})$ as $\Gamma = \sqrt{2} \partial_q\kappa_1(q_s) \alpha_s$ and
the overall effective coupling in the system as  $G = \sqrt{2}  \alpha_s(g_\text{L}+2  g_\text{NL} |\alpha_s|^2)$. Additionally we have defined the effective detuning  $\Delta(q_s)=\Delta-g_\text{L} q_s$. It is worth noting that the nonlinear coupling enters the effective coupling scaled with the mean photon number in the cavity and it can therefore be expected to have a much stronger effect than the linear coupling.

\begin{figure*}
\centering
 \includegraphics[width=.8\linewidth]{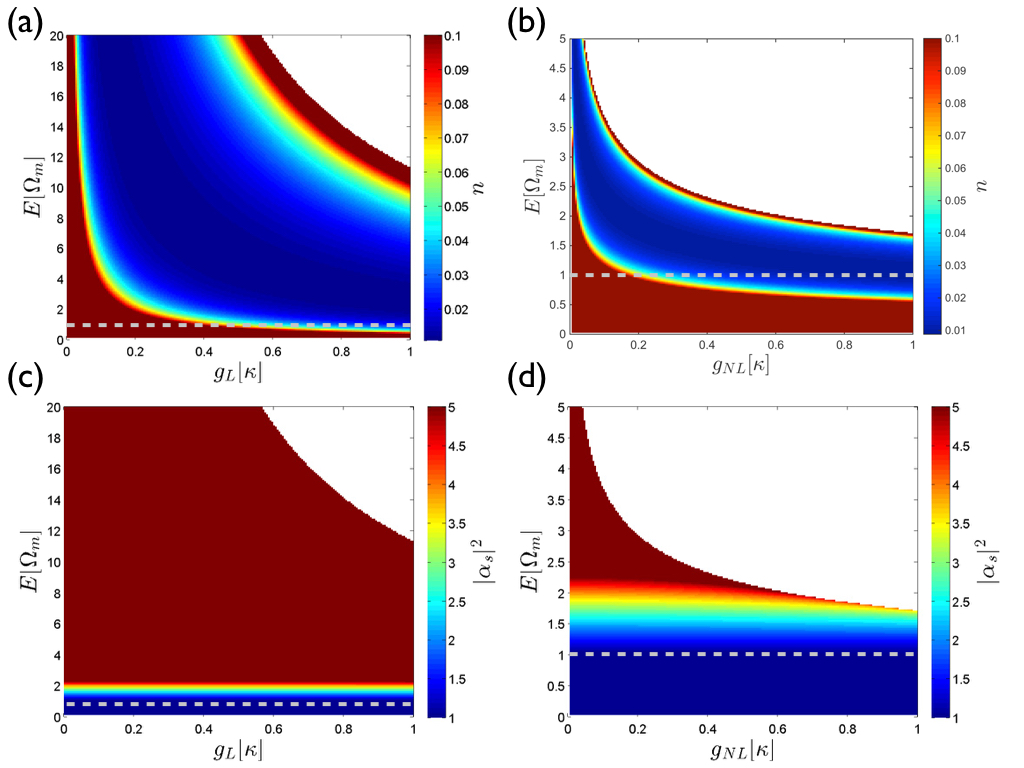}
\caption{(a) Phonon number $n$ as a function of $E$ and $g_\text{L}$ for $g_\text{NL}=0$ and  (b)  as a function of $E$ and $g_\text{NL}$ for $g_{L}=0$.  The color scale is capped at $n=0.1$ for clarity and any larger values are given by a deep red color.  Note the different axis scaling for $E$. (c) Photon number $|\alpha_s|^2$ as a function of $E$ and $g_\text{L}$ for $g_\text{NL}=0$ and (d) as a function of $E$ and $g_\text{NL}$ for $g_{L}=0$.  The color scale is capped below at $|\alpha_s|^2=1$, such that any smaller values are given by a deep blue color, and the point at which $|\alpha_s|^2=1$ is indicated by the gray dashed lines representing the boundary of the applicability of our model. The white areas in all panels correspond to parameter regimes where no stable steady state solutions exist.}
 \label{fig:energydependencies} 
\end{figure*}


The LQLE can be rewritten as the matrix equation of the form 
\begin{equation}
\dot{u(t)}=A u(t)+c(t)
\end{equation}
where 
\begin{align}
u(t) &= \begin{bmatrix}
  \delta \hat{q}(t) \\
  \delta \hat{p}(t) \\
  \delta \hat{X}(t) \\
  \delta \hat{Y}(t) 
\end{bmatrix}
\end{align}
and
\begin{align}
c(t) &= \begin{bmatrix}
  0 \\
  \hat{\xi}(t) +\frac{\Gamma}{\sqrt{2\kappa_1(q_s)}} \hat{Y}_1^{in}(t) \\\
  \sqrt{2\kappa_0}\hat{X}_0^{in}(t)+\sqrt{2\kappa_1(q_s)}\hat{X}_1^{in}(t) \\
 \sqrt{2\kappa_0}\hat{Y}_0^{in}(t)+\sqrt{2\kappa_1(q_s)}\hat{Y}_1^{in}(t)
\end{bmatrix}.
\end{align}
The drift matrix is given by
\begin{equation}
A = \begin{bmatrix}
    0 & \Omega_m & 0 & 0  \\
   -\Omega_m & -\gamma_m & G & 0 \\
     -\Gamma & 0 & -\kappa(q_s) & \Delta(q_s)-2 g_\text{NL} q_s \\
    G & 0 & -\Delta(q_s)+6 g_\text{NL} q_s  & -\kappa(q_s)
\end{bmatrix}
\end{equation}
and can be seen to reduce to the linear drift matrix for $g_\text{NL}=0$ \cite{Biancofiore2011}. The four eigenvalues appear as complex conjugate pairs and give information about the quantum fluctuations in the system. Their real parts corresponds to the cooling (or heating rate) of the membrane and the cavity, which means that the system is only stable if both of these are negative and the system relaxes towards a steady state. The imaginary parts describe the dressed eigenfrequencies of the membrane and the optical field. 
The stationary state is characterized by the covariance matrix 
V, which is determined by the Lyapunov equation
\begin{equation}
AV+VA^T=-D,
\end{equation}
and where D, known as the diffusion matrix, is related to the noise vector $c(t)$ by \cite{Biancofiore2011}
\begin{equation}
D_{lk} \delta(s-s') = [\langle c_k(s) c_l(s') \rangle +\langle c_l(s')c_k(s)\rangle].
\end{equation}
It is explicitly given by \cite{Biancofiore2011} 
\begin{equation}
D = \begin{bmatrix}
    0 & 0 & 0 & 0  \\
  0 & -\gamma_m(2 n_0+1)+\frac{\Gamma^2}{4k_1(q_s)} & 0 & \frac{\Gamma}{2} \\
    0 & 0 & \kappa(q_s) & 0 \\
    0 & \frac{\Gamma}{2} & 0  & \kappa(q_s)
\end{bmatrix}.
\end{equation} 
The mean occupation number for the phonons $n$ can be obtained from the mean energy of the mechanical resonator which is given by
\begin{equation}
U = \frac{\hbar \Omega_m}{2}( \langle \Delta \hat{x}^2 \rangle + \langle \delta \hat{p}^2 \rangle) = \hbar \Omega(n+\frac{1}{2}) = \frac{\hbar \Omega_m}{2}( V_{11} + V_{22}),
\end{equation}
where $V_{11}$ and $V_{22}$ are matrix elements from the covariance matrix.


\begin{figure*}
\centering
 \includegraphics[width=.8\linewidth]{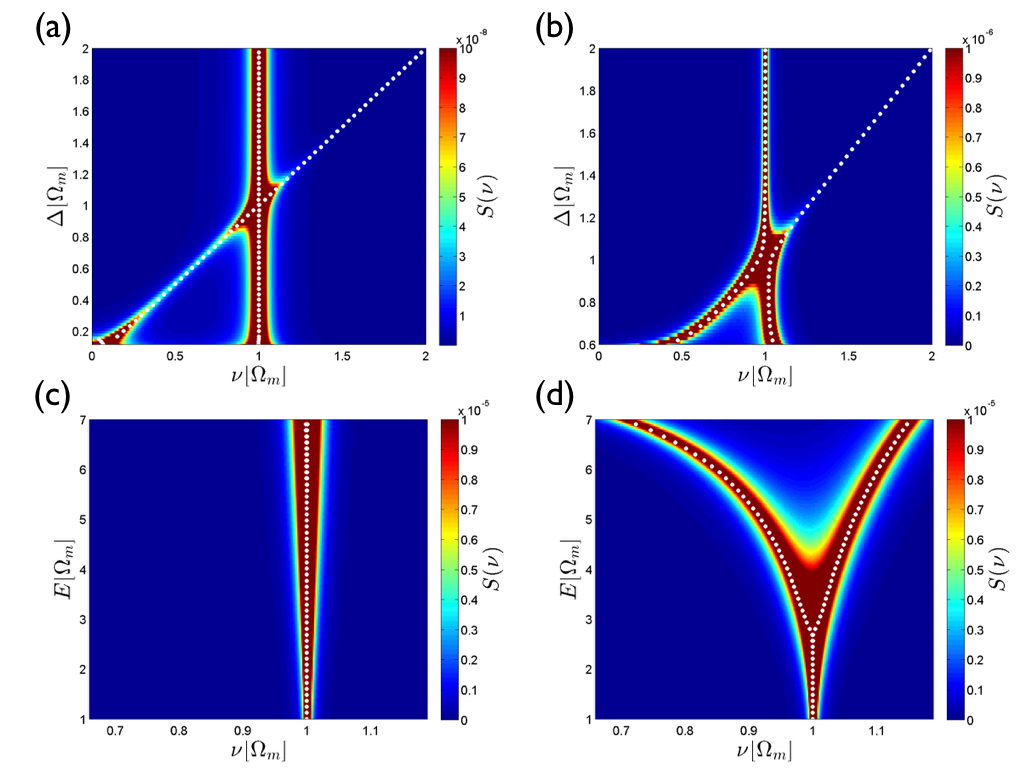}
\caption{(a) Spectrum of a linear system with $g_\text{L}=0.1, g_\text{NL}=0,E=4 \Omega_m$ and (b)  spectrum of the nonlinear system with $g_\text{L}=0.1,g_\text{NL}=0.01 \kappa,E=4 \Omega_m$ both as a function of $\Delta$. 
(c) Spectrum of a linear system with $g_\text{L}=0.1 \kappa,g_\text{NL}=0,\Delta = \Omega_m$ and (d) spectrum of the nonlinear system with $g_\text{L}=0.1,g_\text{NL}=0.01\kappa,\Delta = \Omega_m$ both as a function of $E$. The white dots correspond to the imaginary part of the eigenvalues of the drift matrix $A$.} 
 \label{fig:spectrumplots} 
\end{figure*}

The spectral function for the system is defined as \cite{Kumar2010}
\begin{equation}
S_q(\nu) = \frac{1}{4 \pi} \int d \Omega e^{-i(\nu+\Omega)t} \langle \delta q (\nu) \delta q(\Omega)+\delta q(\Omega)\delta q (\nu) \rangle
\end{equation} 
where $\nu$ is the Fourier space frequency. Taking the Fourier transformation of the LQLE one finds 
\begin{align}
&C_3 \delta \hat{q}(\nu) = G\Big[\sqrt{\kappa_0}\left(\kappa(q_s)-i \nu-i \delta(q_s)\right)\hat{a}_0^{in} \nonumber \\
&+\sqrt{\kappa_0}(\kappa(q_s)-i \nu+i \delta(q_s))\hat{a}_0^{in,\dagger} \nonumber\\
&+\sqrt{\kappa_1(q_s)}\left(\kappa(q_s)-i \nu-i \delta(q_s)-i\frac{C_1 \Gamma}{2 \kappa_1(q_s) G}\right)\hat{a}_1^{in}\nonumber\\
&+\sqrt{\kappa_1(q_s)}\left(\kappa(q_s)-i \nu+i \delta(q_s)+i\frac{C_1 \Gamma}{2 \kappa_1(q_s)G}\right)\hat{a}_1^{in,\dagger}\Big]+C_1 \xi,
\end{align}  
where $C_3 = C_1 C_2-\delta G^2+(\kappa(q_s)-i \nu) \Gamma, C_1 = [\kappa(q_s)-i \omega]^2+\delta'^2$ and $C_2 = \frac{\Omega_m^2+h \omega_m-\nu^2-i \nu \gamma_m}{\Omega_m}$ with $\delta'(q_s)^2=\Delta(q_s)^2+12 g_\text{NL}^2q_s^2-8 \Delta(q_s) g_\text{NL} q_s$ and $\delta(q_s) = \Delta(q_s) - 2 g_\text{NL}q_s$. Using the correlation relations for the noise operators in Fourier space then gives the spectrum
\begin{equation}
S(\nu)=|\chi_{eff}|^2[S_{th}(\nu)+S_{rp}(\nu)+S_{abs}(\nu)],
\end{equation}
where the thermal and radiation pressure spectra are given by
\begin{align}
S_\text{th}(\nu)&=\frac{\gamma_m  \nu}{\Omega_m} \left[1+\coth\left(\frac{\hbar \nu}{2 k_b T_0}\right)\right],\\
S_\text{rp}(\nu) &= \frac{2 G^2 \kappa(q_s)[\kappa(q_s)^2+\nu^2+\delta(q_s)^2]}{(\kappa(q_s)^2-\nu^2+\delta'(q_s)^2)^2+4  \nu^2 \kappa(q_s)^2},
\end{align}
and we have an extra noise spectrum associated with membrane absorption 
\begin{equation}
S_\text{abs}(\nu) = \frac{\Gamma^2}{2 \kappa_1(x_s)}+\frac{2 G \Gamma \delta(q_s)[\kappa(q_s)^2-\nu^2+\delta'(q_s)^2]}{(\kappa(q_s)^2-\nu^2+\delta'(q_s)^2)^2+4  \nu^2 \kappa(q_s)^2}.
\end{equation}
Here $|\chi_\text{eff}|^2$ us the effective susceptibility and is given by 
\begin{align}
& \chi_\text{eff}^{-1}=\frac{1}{\Omega_m}[\Omega_\text{eff}^2-\nu^2]-i\nu \Gamma_\text{eff},\\
& \Omega_\text{eff}^2 = \Omega_m^2 -\Omega_m \frac{[\delta(q_s) G^2(\kappa(q_s)^2-\nu^2+\delta'^2)}{ (\kappa(q_s)^2-\nu^2+\delta'(q_s)^2)^2+4  \nu^2 \kappa(q_s)^2} \nonumber \\
&-\Omega_m \frac{-G \kappa(q_s) \Gamma(\kappa(q_s)^2+\nu^2+\delta'(q_s)^2)}{ (\kappa(q_s)^2-\nu^2+\delta'(q_s)^2)^2+4  \nu^2 \kappa(q_s)^2}, \\
& \Gamma_\text{eff} =\frac{ \gamma_m}{\Omega_m}+\frac{2\delta(q_s) G^2 \kappa(q_s)- G\Gamma(\kappa(q_s)^2+\nu^2-\delta'(q_s)^2)}{ (\kappa(q_s)^2-\nu^2+\delta'(q_s)^2)^2+4  \nu^2 \kappa(q_s)^2}. 
\end{align}
Again we note that the spectrum reduces to the linear one discussed in Ref. \cite{Biancofiore2011} if we assume that $g_\text{NL}=0$.
\section{Numerical results}
To stress the effects stemming from the nonlinear nature of the coupling, we focus in the following on results that show significant differences to the linear situation. Furthermore, for the clearest comparison we initially make the approximation that $\kappa_1(\hat{q})=\kappa_1$, which reduces the problem to the standard optomechanical setup. This approximation is also consistent with existing experimental data for a linear membrane \cite{Karuza2012}. 
In general the behavior of our system can be distinguished into two categories.
In the first one, $g_\text{L}$ and $g_\text{NL}$ have the same sign and thus enhance each other. In this regime the nonlinear coupling leads to a very strong effective coupling but no qualitative differences from the system where only a linear coupling is present. In the second category $g_\text{L}$ and $g_\text{NL}$ have opposite signs, i.e., one of them is attractive and the other is repulsive. Here a parameter regime exists for which additional steady-state solutions, that are not present for the linear system, can be found. These arise from the seventh-order polynomial describing the steady state of the nonlinear system. In our calculations we use a mechanical frequency of $\frac{\Omega_m}{2 \pi} =356.6\; {\rm kHz}$ and a loss rate of $\kappa/2\pi= 77\; {\rm kHz}$ corresponding to the experimental values in Ref. \cite{Karuza2012}. We generally employ a mechanical dissipation given by $\gamma_m = 0.01 \kappa$ and choose a temperature which gives an initial phonon number of $n_0=1$.

\subsection{Nonlinear enhancement}

\begin{figure}
\centering
\includegraphics[width=.9\linewidth]{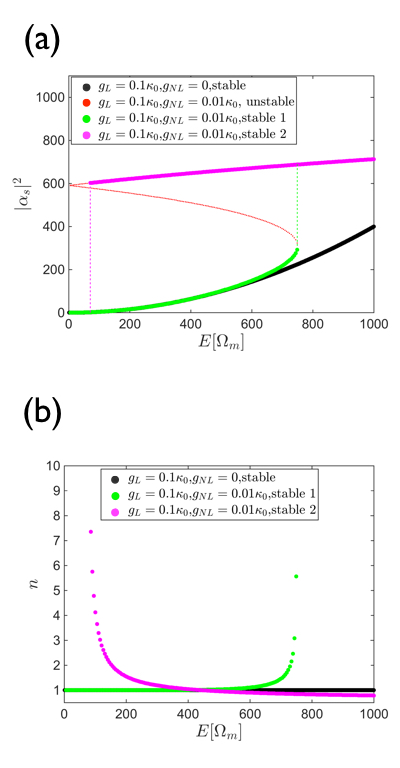}
\caption{(a) Photon number $|\alpha_s|^2$ and (b) phonon number $n$ as a function of $E$ for both the linear and nonlinear case. The thin red (dotted) line corresponds to the unstable solutions, while the thicker lines are the various stable solutions.} 
 \label{fig:3sol} 
\end{figure}

The phonon number as a function of the input energy of the laser and the linear and nonlinear couplings, respectively, for $\Delta = \Omega_m$ is shown in Figs. \ref{fig:energydependencies}(a) \ref{fig:energydependencies}(b).  This corresponds to the  resolved sideband regime where optimal cooling can be expected. One can see that for any value of $g_\text{L}$ or $g_\text{NL}$ there is a value of $E$ for which the same maximal cooling ($n \approx 0.01$) is obtained and this value reduces as the coupling strength is increased. In fact, the nonlinear coupling simply enhances this behavior that is already present in the linear case, reducing the value of $E$ required to achieve the same amount of cooling as in the linear case.  Additionally one can see a rise in temperature as $E$ increases beyond the value for maximal cooling, until the solutions become unstable (white areas in the figures).  The photon numbers for the same parameter ranges are plotted in Figs. \ref{fig:energydependencies}(c) and \ref{fig:energydependencies}(d). One can see that they generally increase with energy as one would expect; however, it is more relevant to confirm whether the photon number stays above 1 in the regime of relevance, i.e., the regime of maximal cooling, since our model is otherwise inapplicable. In fact, for very small values of $E$ it drops below 1 and to get maximal cooling for large values of $g_\text{NL}$, values of $E$ approaching this inapplicable regime must be chosen. For the most part, however, this limitation does not pose a problem. Situations where both $(g_\text{L}$ and $g_\text{NL})$ are finite behave similarly to the two limiting cases discussed here. 
\begin{figure*}
\centering
\includegraphics[width=.8\linewidth]{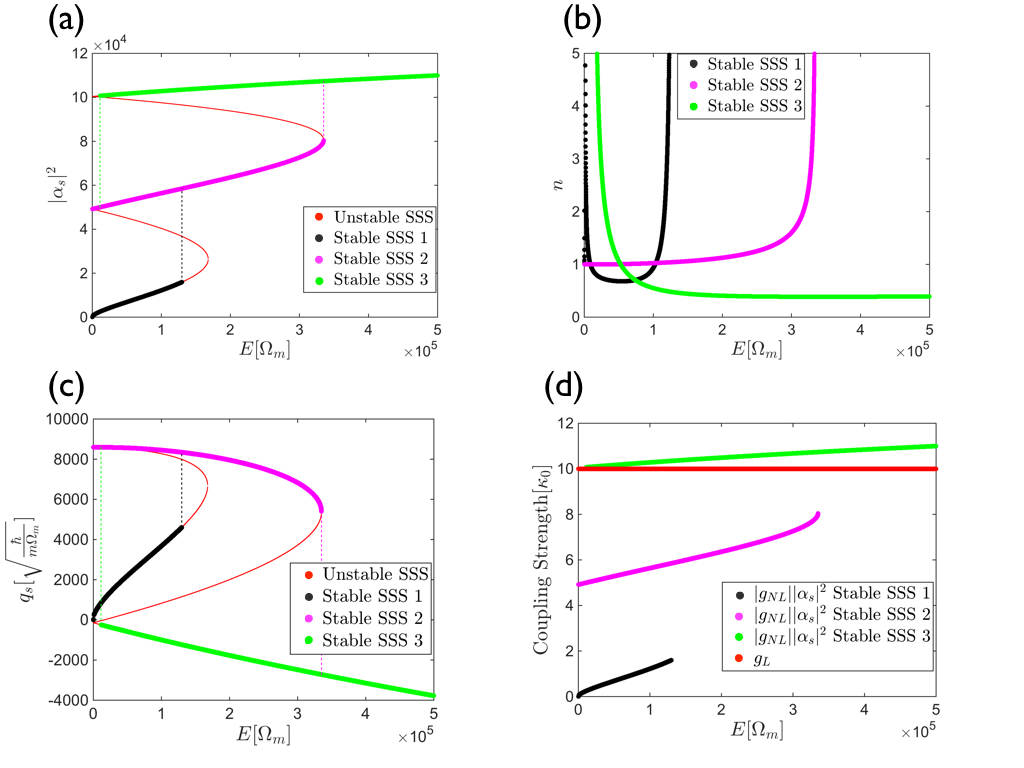}
\caption{(a) Photon number $|\alpha_s|^2$, (b) phonon number $n$, and (c) position of the membrane $q_s$ as a function of $E$. The thin red line corresponds to the unstable solutions, while the thicker lines are the three stable solutions. (d) $g_\text{NL} |\alpha_s|^2$ and $g_\text{L}$ as a function of $E$.} 
 \label{fig:multistability} 
\end{figure*}

To investigate the existence of a strong-coupling regime at small laser intensities, we will look next at the spectrum $S(\nu)$ for $(g_\text{L},g_\text{NL})=(0.1 \kappa,0)$ and $(g_\text{L},g_\text{NL})=(0.1 \kappa,0.01 \kappa)$. In both cases we have a fairly weak linear coupling and therefore do not expect the first case to be in the strong-coupling regime. The spectrum as a function of the detuning $\Delta$ at $E=4 \Omega_m$ is plotted in Figs. \ref{fig:spectrumplots}(a) and \ref{fig:spectrumplots}(b), and one can see that without the nonlinear coupling the system is indeed in the weak-coupling regime (no splitting is visible).  However, adding a comparatively small amount of nonlinear coupling, $g_\text{NL}=0.1 g_\text{L}$, brings the system into the strong-coupling regime, signified by the avoided crossing of the dressed eigenfrequencies. This is quite remarkable as it allows for strong coupling at much smaller laser energies by adding just a small nonlinearity to the system.  The spectrum as a function of the energy $E$ at $\Delta = \Omega_m$, is plotted in Figs. \ref{fig:spectrumplots}(c) and \ref{fig:spectrumplots}(d), where the normal-mode splitting for the nonlinear coupling $g_\text{NL}=0.1 g_\text{L}$ is distinctly visible and clearly absent for the linear case at these energies.  The figures also shows that our results are self-consistent as the imaginary part of the eigenvalues of the drift matrix $A$ (dotted white lines) correspond to the two peak positions in frequency space.

Another situation where the dramatic effect of a small nonlinearity becomes visible is shown in Fig.~\ref{fig:3sol}(a). Here the photon number is given as a function of $E$  for $(g_\text{L},g_\text{NL})=(0.1 \kappa,0)$ and $(g_\text{L},g_\text{NL})=(0.1 \kappa,0.01 \kappa)$ at $\Delta = 50 \Omega_m$.
The linear system can be seen to have only one solution in the displayed energy range, but adding the small nonlinearity leads to optical bistability as evidenced by the characteristic S-shaped curve. Starting at $E=0$ and slowly increasing the energy the system moves along the first stable solution with the smallest number of photons. When this solution becomes unstable there is a first order transition to the second stable solution which has the largest number of photons. If the energy is then decreased the system stays in this second steady state until it becomes unstable at which point a first-order transition to the first steady state takes place. This is the characteristic signature of optical hysteresis. In Fig.~\ref{fig:3sol}(b) the corresponding phonon numbers are plotted. The temperature (phonon number) of the solutions are fairly constant in the stable regimes but rise rapidly as a solution becomes unstable.

It is worth stressing again that this optical bistability appears at quite  low energies, because the nonlinearity leads to strong effective coupling. To see bistability with linear coupling only,  much larger energies would be needed.  
 
\subsection{Optical multistability}
While we have shown above that  choosing the linear and the nonlinear coupling strength to have the same sign leads to mostly an enhancement of the effects already present in the linear case, qualitatively new effects can appear when they have opposite signs. 
For this we consider in the following $(g_\text{L},g_\text{NL})=(10 \kappa,-10^{-4} \kappa)$ and $\Delta = 50 \Omega_m$. The reason for the large difference between the linear and the nonlinear coupling strength is that if the couplings were chosen to be of comparable size, the nonlinear coupling will dominate the system as it enters the effective coupling scaled with the photon number, i.e., $G = \sqrt{2}  \alpha_s(g_\text{L}+2  g_\text{NL} |\alpha_s|^2)$. Therefore, in order to engineer competition between the two forces, values for which the contribution of the nonlinear coupling $g_{\text{NL}} |\alpha_s|^2$ and the linear coupling $g_\text{L}$ are comparable must be chosen. In Fig.~\ref{fig:multistability}  we show the behavior of $|\alpha_s|^2,n,q_s,$ and $g_{\text{NL}} |\alpha_s^2$. Five steady-state solutions can be found, but only three of them are stable. To ease the discussion we will denote the three stable branches as 1, 2, and 3, where 1 corresponds to the stable solution with the smallest number of photons, 2 corresponds to the middle solution, while 3 corresponds to the solution with largest photon number [see Fig.~\ref{fig:multistability}(a)]. For increasing $E$, starting from $E=0$  the system will be in a state on the first branch, jump to the second branch at energies where the first branch is no longer stable, before finally moving to the third branch. Looking at the position of the membrane for these states, one can see from Fig.~\ref{fig:multistability}(c) that it moves in the positive direction along the first branch, then changes direction and moves in the negative direction along the second branch, and, finally, jumps to a negative value moving in the negative direction when the system enters the third branch. This means that as the system jumps from the second to the third branch, the position of the membrane should jump from being displaced in the positive direction to being displaced in the negative one. Since $q_s =(g_\text{L}+g_\text{NL}|\alpha_s|^2)\frac{|\alpha_s|^2}{\Omega_m}$ one can understand this behavior by looking at the size of $g_\text{L}$ and $g_\text{NL} |\alpha_s|^2$, which is plotted in Fig.~\ref{fig:multistability}(d). Along the first branch $g_\text{L}$, which is positive, is dominant and so the membrane is pushed in the positive direction. Along the second branch $g_\text{NL} |\alpha_s|^2$, which depends on the photon number, becomes comparable with $g_\text{L}$ and though $q_s$ stays positive the effective coupling $G$ becomes negative. Increasing values of $E$  mean an effective decrease in the (positive) position $q_s$. On the third branch $g_\text{NL} |\alpha_s|^2$ is always much larger than $g_\text{L}$ and therefore the membrane has a negative effective coupling, while also being at a negative value in position space. This means that the system moves between attractive and repulsive regimes for the interaction between the membrane and the cavity field as it jumps between stable solutions. Finally, looking at the phonon number in Fig.~\ref{fig:multistability}(b), one can see that the temperature (phonon number) of the second branch is considerably higher than that of the other two, with the third branch having the lowest temperature. 

\begin{figure}
\centering
 \includegraphics[width=.9\linewidth]{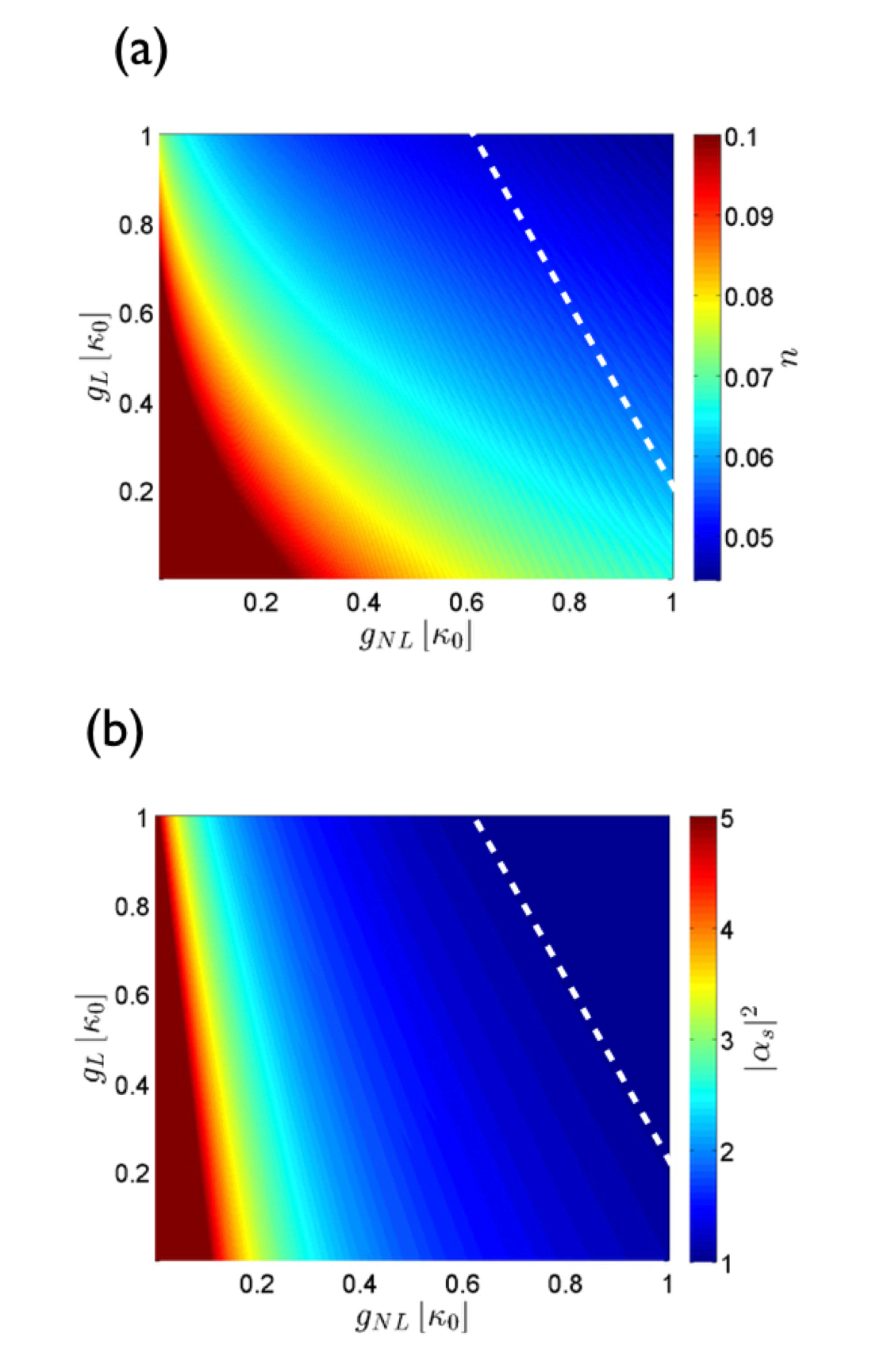}
\caption{(a) Phonon number $n$ as a function of $g_\text{L}$  and $g_\text{NL}$, where $E$ has been chosen so as to minimize the temperature for each pair.  (b) Photon number $|\alpha_s|^2$ as a function of $g_\text{L}$  and $g_\text{NL}$ for the same steady-state solutions. The dashed white lines show where the photon number is equal to unity.} 
 \label{fig:g_Nlvsg_L} 
\end{figure}

\subsection{Effect of position-dependent absorption}
Finally we consider the case of letting the membrane absorption depend on the position, i.e., $\kappa_1(\hat{q})=\kappa_\text{L} \hat{q}$ with $\kappa_\text{L} = 0.0130 \kappa_0$ and $\kappa_0 = 77 {\rm kHz}$, and study how this affects the cooling in the resolved sideband regime, where $\Delta = \Omega_m$. 
For this, the phonon number $n$ and the photon number $|\alpha_s|^2$  as a function of $g_\text{NL}$ and $g_{L}$ are shown in Fig.~\ref{fig:g_Nlvsg_L}, where the energy  at each point has been chosen so as to minimize $n$. 
One can see that the position dependent absorption coefficient leads to overall worse cooling [compare with $n \approx 0.01$ from Figs. \ref{fig:energydependencies} (a) and \ref{fig:energydependencies} (b)], which is to be expected.  However, by increasing the coupling this trend can be counteracted, leading to more efficient cooling for a more strongly coupled system. As can be seen from Fig.~\ref{fig:g_Nlvsg_L} this favors the nonlinear coupling, which is always stronger than the linear coupling. Therefore in this case the nonlinear coupling can lead to more efficient cooling. Looking at the photon number however, one must be wary as the energy for which optimal cooling is obtained corresponds to photon numbers on the order of one for very strongly coupled systems. While this means that the basic assumption for our model breaks down, we find enhanced cooling for the majority of the investigated ranges without encountering this problem.

\section{Conclusion}
Applying the Langevin formalism we have derived the classical steady-state solutions and the spectrum of quantum fluctuations for a generic optomechanical system including a nonlinear position-modulated self-Kerr optomechanical interaction. We find that the the effective nonlinear coupling scales with the square of the photon number, which implies that even for a small nonlinear coupling the system enters the strong-coupling regime. By analyzing the obtained solutions numerically this is confirmed as a small nonlinear coupling leads to normal-mode splitting and an avoided crossing in the spectrum, which is associated with the strong-coupling regime. This also leads to lower powers being required for motional cooling and bistability compared to systems where only a linear coupling is present. Furthermore, we find that the addition of a weak nonlinear coupling with the opposite sign of the linear coupling leads to three stable solutions, who can all be occupied as a function of the energy. Finally, we find that in the case of position-dependent absorption the nonlinear coupling can help counteract the degradation of cooling previously predicted in this regime. All these effects are consistent with a strong effective nonlinear coupling, even at small coupling strengths $g_\text{NL}$. This means that engineering an effective nonlinear system experimentally is easier than what would initially be assumed, as only a small nonlinearity is required to observe dramatic effects. 

Since it is known that nonlinear systems can support self-sustained oscillations  and researchers have investigated these in the case of  optomechanical systems possessing motional Kerr optical nonlinearities \cite{Mar2006, Lud2008}, we expect that with nonlinear optical coupling the behavior of such self-sustained oscillations may be even richer and an interesting topic for future investigations.  Some possible avenues of experimental realization for such systems are cavity polaritons \cite{Bobrovska2016} and atoms trapped near fibre tapers \cite{Kato2015, Polzik2016}, but the most promising one is utilizing artificial multilevel cooper pair box molecules in a microwave cavity \cite{Rebic2009}, which we discuss in the appendix of this paper.

\section*{Acknowledgment}
This  project  was  supported  by  the  Okinawa  Institute  of Science and Technology Graduate University and the Australian Research Council Centre of Excellence in Engineered Quantum Systems.

\section*{Appendix: Obtaining the nonlinear coupling constant from an electromechanical microwave setup}
We consider the artificial multilevel cooper pair box molecule interacting with a microwave coplanar resonator, as described in Ref. \cite{Rebic2009}. 
The general idea is to engineer an artificial atom as a four-level N-system, and to create a Kerr nonlinearity using a scheme similar to the EIT scheme for atoms  introduced by Schmidt and Immamoglu \cite{Schmidt1996}. For this the four-level effective Hamiltonian in terms of the electric properties of the superconducting circuit is derived, which gives explicit descriptions of the nonlinear coupling coefficients. Since the mutual capacitance between the cooper pair boxes depends on the distance between them, the capacitance becomes position-dependent if one of the boxes is coupled to a moving membrane. This leads to a parametric dependence on the position $\hat{x}$ of the form
\begin{equation}
C_m(\hat{x})=\frac{\epsilon_r \epsilon_0 A}{d_0-\hat{x}},
\end{equation} 
where $\epsilon_0$ is the vacuum permittivity, $\epsilon_r $ is the permittivity of the material between the plates, $A$ is the area of both plates ,and $d_0$ is the initial distance between the plates.
This means that any quantity, including $\eta$ which depends on the capacitance $C_m$, becomes position-dependent as well. In Ref.~\cite{Rebic2009} it is shown that the nonlinear coefficient $\eta$ in the regime where $(\frac{g_1}{\Omega_C})^2 \ll 1$ is  given by    
\begin{equation}
\eta(\hat{x}) = \frac{g_1^2}{\Omega_C^2}\left( \frac{g_2^2 \Delta(\hat{x})}{\gamma_{43}^2+\Delta(\hat{x})}-\frac{g_1^2 \delta(\hat{x})}{ (\gamma_{21}+\gamma_{23})^2+\delta(\hat{x})^2} \right),
\end{equation}  
where $\Delta$ and $\delta$ are the detunings between the cavity frequency $\omega_c$ and the fourth and second energy states $| 4 \rangle$ and $| 2 \rangle$, respectively. The $\gamma_{ij}$ are the spontaneous decay rates from $| i \rangle$ to $|j \rangle$, $g_1$ and $g_2$ are the coupling strengths between $| 1 \rangle, | 2 \rangle$ and $| 3 \rangle, | 4 \rangle$, respectively, and, finally, $\Omega_C$ is the coupling between $| 2 \rangle$ and $ | 4 \rangle$   (see  Fig.~1 in Ref. \cite{Rebic2009}).  The position-dependence is contained in the two detunings $\delta$ and $\Delta$ as these depend on $C_m$.  Analytic expressions for the these detunings can be found at the coresonance point where they become equal, $\delta = \Delta$, and are given by \cite{Rebic2009}
\begin{equation}
\Delta(\hat{x})=\sqrt{J(\hat{x})^2+4 \omega_x^2}+J(\hat{x})-\omega_c(\hat{x})
\end{equation}
where 
\begin{equation}
\hbar J(\hat{x}) = \frac{0.5 (2 e)^2C_m(\hat{q})}{4(k_1 C_m(\hat{x})+k_2)},
\end{equation}
with $k_1 = C_{\Sigma 1}+C_{\Sigma 2}$ and $k_2 = C_{\Sigma 1}C_{\Sigma 2}$. Here $e$ is the elementary charge, while $C_{\Sigma 1},C_{\Sigma 2}$ are capacitances coming from different parts of the circuit (see. Fig.~2 in Ref. \cite{Rebic2009} for a diagram of the circuit). Depending on the setup, the resonant frequency $\omega_c(\hat{x})$ can also be dependent on the the position coordinate (and in fact has to be if we wish to engineer a linear coupling). In order to simplify our estimate of the nonlinear coefficient, we will also assume that $g_1=g_2=g$ and $\gamma_{21}=\gamma_{23}=\gamma_{43}=\gamma$ (similarly to Ref. \cite{Rebic2009}), which leads to
\begin{equation}
\eta(\hat{q}) = \Delta(\hat{x})\frac{g^4}{\Omega_C^2}\left( \frac{1}{\gamma^2+\Delta(\hat{x})^2}-\frac{1}{ 4\gamma^2+\Delta(\hat{x})^2} \right)
\label{eq:etaofq}
\end{equation} 
To find the nonlinear coefficient we then need to evaluate the derivative of $\eta$ at $\hat{x}=0$ (see Sec. II.A), which we can obtain analytically by simply taking the derivative of Eq. (\ref{eq:etaofq}). In order to evaluate at $\hat{x}=0$ we require
\begin{equation}
C_m(0)=C_0= \frac{\epsilon_r \epsilon_0 A}{d_0} ,
\end{equation}
and 
\begin{equation}
\frac{d C_m(\hat{x})}{d \hat{x}}\bigg|_{\hat{x}=0}= \frac{\epsilon_r \epsilon_0 A}{d_0^2}=\frac{C_0}{d_0}. 
\end{equation}
To estimate the obtainable sizes of $G_\text{NL}$ and $\eta_0$ we need to input physically realistic values for the different parameters.  We use estimates similar to those in Ref. \cite{Rebic2009},  $g / 2 \pi= 300 \kappa, \kappa = 1 \text{ MHz}, \gamma= 10 \text{ MHz}, \Omega_C/ 2 \pi = 0.9478 \text{ GHz}$. Additionally, we model the mutual capacitance, using values corresponding to those in the experimental setup from Ref. \cite{Teufel2011} with  $C_0=940 \text{ fF},d_0 = 50 \text{ nm}$ and use a similar cavity frequency $\omega_c/2 \pi  =  7.54 \text{ GHz}$ as well as assuming a linear coupling equivalent to what they obtain, $G_\text{L}/ 2 \pi= 49\text{ MHz/nm}$. Finally, we use the values  $C_{\Sigma 1}=C_{\Sigma 2} = 4 \text{ fF}$ \cite{Gywat2006}. It turns out that $\omega_x$ is a useful knob for changing $G_\text{NL}$, while keeping the other values constant, 
and one can get a large range of values for the nonlinear coefficients: For example, for $\omega_x / 2 \pi= 3.3595  \text{ GHz}$  we get $G_\text{NL}/ 2 \pi= 95.3  \text{ MHz/nm}, \eta_0 / 2 \pi= -0.751\text{ MHz}$, while $\omega_x / 2 \pi= 3.34 \text{ GHz}$ gives $G_\text{NL}/ 2 \pi =  0.637 \text{ kHz/nm},\eta_0/ 2 \pi = 0.1001\text{ kHz}$. The nonlinear coupling can be tuned over several orders of magnitude and while much smaller values than the ones depicted above are easily obtainable, the main takeaway is that the nonlinear coupling can be made comparable in size to the linear coupling.  To get the nondimensional coupling coefficients we multiply by the zero-point position from the experimental setup in Ref. \cite{Teufel2011}  $x_{zp}=4.1 \text{ fm}$ which for the first case above corresponds to $g_\text{L} = 1.26 \text{ kHz}, g_\text{NL} = 2.46 \text{ kHz}$. Finally, we note that it is possible to place a second artificial molecule within the cavity, which does not couple to the position element, in such a way that the term $\eta_0$ is canceled out. This is important, as it allows the nonlinear interaction term to be engineered without the photon blockade effect associated with the $\hat{H}_\text{kerr}$ term.  In order to engineer a value $\eta$ of the same size with the opposite sign, the easiest way is to simply engineer a detuning with the opposite sign, i.e.,  $\Delta_\text{stationary} = -\Delta_\text{moving}(0)$, keeping the other parameters constant. By allowing the parameters  $\delta, \Delta, \gamma_{21},\gamma_{23},\gamma_{43}$ to vary as well there are, however, many more ways to engineer an artificial molecule which cancels  $\eta_0$. 

\FloatBarrier

\end{document}